\documentclass[amsmath,amssymb,nofootinbib]{revtex4}

\usepackage{graphicx}
\usepackage{dcolumn}
\usepackage{bm}

\begin{document}


\title{Particle-number conservation\\ within 
self-consistent random-phase approximation\footnote{Accepted in 
Physical Review C}}

\author{Nguyen Dinh Dang}
\altaffiliation[E-mail address: ]
{dang@riken.jp}
\affiliation{%
Cyclotron Center, RIKEN, 2-1 Hirosawa, Wako, 351-0198 
 Saitama, Japan}%

\begin{abstract}
The self-consistent random-phase approximation (SCRPA) is reexamined 
within a multilevel-pairing model with double degeneracy. 
It is shown that the expressions for 
occupation numbers used in the original version of SCRPA violate
the particle number for non-symmetric particle-hole ($ph$) spectra. 
A renormalization is introduced to restore the 
particle number, which leads to the expressions of occupation numbers 
similar to those derived by Hara et al. for the 
$ph$ case. The results of calculations within the 
$ph$ symmetric case show that this 
number-conserving SCRPA yields the energies of the ground state and 
first excited state of the system with $\Omega+2$ particles relative 
to the ground state of the system with $\Omega$ particles
in close agreement with those obtained within the original SCRPA. However
it gives a slightly larger correlation energy. 
\end{abstract}
\pacs{21.60.Jz, 21.60.Cs, 21.90.+f}

\maketitle
\section{Introduction}
The random-phase approximation (RPA) has been a powerful tool
in the theoretical study of many-body systems such as atomic nuclei.
An essential ingredient of the RPA is the use of the quasiboson 
approximation (QBA), which considers fermion pairs as boson
operators, just neglecting the Pauli principle between them.
Within the QBA a set of linear equations, which is usually called the RPA 
equation, is derived, which makes computationally demanding 
problems become tractable. However, because of the violation of 
Pauli principle within the QBA, the RPA equation breaks down 
at a certain critical value of the interaction's parameter, 
where the RPA yields imaginary solutions.

Several approaches were developed to remove this inconsistency of the 
RPA. One of the popular ones is the renormalized RPA 
(RRPA)~\cite{Hara,Rowe,Schuck,CaDaSa}.
The RRPA includes in the expectation 
value over the ground state the contribution of the diagonal
elements of the commutator between two fermion-pair operators. In 
this way it takes the Pauli principle into account approximately. 
This includes the so-called ground-state correlations beyond 
RPA, which  eventually renormalize the interaction in such a way that the collapse of 
RPA is avoided. However, the tests carried out within 
exactly solvable models also showed that 
there is still a large discrepancy between 
the solution obtained within the RRPA and the exact one 
beyond the RPA collapsing point (See Ref. \cite{CaDaSa} e.g.).

The situation has been significantly improved recently within the
self-consistent RPA (SCRPA)~\cite{Duk,Dukelsky,Hirsch} due to the inclusion of 
screening corrections in the SCRPA equation. These screening 
corrections are in fact the expectation values of the products of 
two fermion pairs in the correlated 
ground state. As the result the sign of the interaction is reversed
so that, within a particle-hole ($ph$) symmetric 
multilevel-pairing model with double degeneracy (the so-called 
picket-fence model), the SCRPA yields the solutions very close to the exact ones
for the correlation energy of the system with $\Omega$ particles,
as well as the energy of the first excited 
state of the system with $\Omega+2$ particles~\cite{Dukelsky,Hirsch}.

Realistic nuclear single-particle spectra are in general  
$ph$ non-symmetric, which means that the particle-particle ($pp$)
submatrix $A$ and hole-hole ($hh$) submatrix $C$ of the $pp$-RPA 
equation do not have the same dimension. The asymmetry is 
particularly strong, e.g., in light neutron-rich nuclei
~\cite{VinhMau}, 
for which the effect due to Pauli principle cannot be 
neglected. 
It is, therefore, worthwhile to reexamine carefully the SCRPA
before applying it to realistic nuclei.

The present paper employs the same picket-fence model, 
which was used to tested the validity of the SCRPA in Refs. 
\cite{Dukelsky,Hirsch}. It will be shown that, in the general 
$ph$ non-symmetric case, using its original expressions 
of ground-state correlation factors, the SCRPA violates the particle number.
A simple and consistent way to restore the particle number will be introduced and the 
consequences will be discussed.

The paper is organized as follows. The outline of the SCRPA for the 
picket-fence model is presented in Sec. \ref{outline}. The 
violation of particle number within the SCRPA for $ph$ non-symmetric 
case and the construction of a number-conserving SCRPA
are discussed in Sec. \ref{violationN}. The results of numerical 
calculations are analyzed in Sec. \ref{numerical}. The paper is 
summarized in the last section, where conclusions are drawn.
\section{SCRPA equation for the picket-fence model}
\label{outline}
The detail derivation of self-consistent $pp$ RPA equation, 
which is simply called SCRPA equation hereafter 
has been described Refs.
\cite{Duk,Dukelsky,Hirsch}. The present section recuperates only 
the brief outline of the SCRPA for the picket-fence model, which is 
needed for the discussion in this paper.
\subsection{Model Hamiltonian}
\label{model}
The picket-fence model consists of $\Omega$ two-fold equidistant levels interacting 
via a pairing force with a constant parameter $G$. 
The model Hamiltonian is written as
\begin{equation}
H=\sum_{i=1}^{\Omega}(\epsilon_{i}-\lambda)N_{i}-G\sum_{i,j=1}^{\Omega}
P_{i}^{\dagger}P_{j}~,
\label{H}
\end{equation}
where the particle-number operator $N_{i}$ and pairing operators 
$P_{i}^{\dagger}$, $P_{i}$ are given as
\begin{equation}
N_{i}=c_{i}^{\dagger}c_{i}+c_{-i}^{\dagger}c_{-i}~, 
\hspace{5mm} 
P_{i}^{\dagger}=c_{i}^{\dagger}c_{-i}^{\dagger}~,\hspace{5mm} 
P_{i}=(P_{i}^{\dagger})^{\dagger}~.
\label{N&P}
\end{equation}
The exact commutation relations between the operators $N_{i}$, 
$P_{i}^{\dagger}$, and $P_{i}$ are
\begin{equation}
[P_{i}, P_{j}^{\dagger}]=\delta_{ij}(1-N_{i})~,
\label{[PP]}
\end{equation}
\begin{equation}
[N_{i}, P_{j}^{\dagger}]=2\delta_{ij}P_{j}^{\dagger},~\hspace{5mm} [N_{i}, 
P_{j}]=-2\delta_{ij}P_{j}~.
\label{[NP]}
\end{equation}
The single-particle energies take the values 
$\epsilon_{i}=i\epsilon$ with $i$ running over all $\Omega$ levels.
The original version of the SCRPA in Refs. \cite{Dukelsky,Hirsch} 
was applied only to a $ph$-symmetric spectrum, where there are as many 
particles as levels (half filling). This means that,  
in the absence of interaction ($G=$0), the lowest $\Omega_{h}=\Omega/2$ levels are occupied
with $N=\Omega$ particles (two particles in each level). However, in general, the number 
$\Omega_{h}$ of 
hole levels is not necessary to be the same as the number 
$\Omega_{p}$ of particle 
levels, i.e. $\Omega_{h}\neq\Omega_{p}\neq\Omega/2$.
Numerating particle ($p$) and hole ($h$) levels 
from the levels closest to the Fermi surface, 
the particle and hole energies are equal to 
$\epsilon_{p}=\epsilon(\Omega_{h}+p)$ and 
$\epsilon_{h}=\epsilon(\Omega_{h}-h+1)$, 
respectively, with 
$h$ indices running from 1 to $\Omega_{h}$, and $p$ indices
running from 1 to $\Omega_{p}=\Omega-\Omega_{h}$.
The Fermi level $\lambda$ is defined in the middle of the
the first $h$ and the first $p$ levels, i.e. 
\begin{equation}
    \lambda=\epsilon\big(\Omega_{h}+\frac{1}{2}\big)-\frac{G}{2}~.
    \label{lambda}
\end{equation}
Therefore, in the $ph$ symmetric case 
($\Omega_{p}=\Omega_{h}=\Omega/2$), the Fermi level 
$\lambda=[\epsilon(\Omega+1)-G]/2$
is located in the middle of the single-particle spectrum.

Using the notation of Ref. \cite{Hirsch}
\begin{equation}
M_{p}=N_{p}~,\hspace{5mm} M_{h}=2-N_{h}~,\hspace{5mm} 
Q^{\dagger}_{p}=P^{\dagger}_{p}~,\hspace{5mm} Q_{h}=-P^{\dagger}_{h}~,
\label{M&Q}
\end{equation}
and also introducing the ground-state correlation operators 
$D_{p}$ and $D_{h}$
\begin{equation}
    D_{p}=1-M_{p}=1-N_{p}~,\hspace{5mm} D_{h}=1-M_{h}=N_{h}-1~,
    \label{D}
\end{equation}
the exact commutation relations (\ref{[PP]}) and 
(\ref{[NP]}) can be transformed into
\begin{equation}
[Q_{p}, Q_{p'}^{\dagger}]=\delta_{pp'}D_{p}~,\hspace{5mm} 
[Q_{h}, Q_{h'}^{\dagger}]=\delta_{hh'}D_{h}~,
\label{[QQ]}
\end{equation}
\begin{equation}
[M_{i},Q_{j}^{\dagger}]=2\delta_{ij}Q_{j}^{\dagger}~,\hspace{5mm} 
[M_{i},Q_{j}]=-2\delta_{ij}Q_{j}~.
\label{[MQ]}
\end{equation}
Using the definition (\ref{lambda}) of the Fermi energy $\lambda$ 
together with 
notations (\ref{M&Q}) and (\ref{D}), Hamiltonian
(\ref{H}) is rewritten in the following form
\[
H=-\epsilon\Omega_{h}^{2}
+\sum_{p=1}^{\Omega_{p}}\big[\epsilon(p-\frac{1}{2})+\frac{G}{2}\big]M_{p}
+\sum_{h=1}^{\Omega_{h}}\big[\epsilon(h-\frac{1}{2})+\frac{G}{2}\big]M_{h}
\]
\begin{equation}
-G\sum_{pp'}Q_{p}^{\dagger}Q_{p'}-G\sum_{hh'}Q_{h}^{\dagger}Q_{h'}
+G\sum_{ph}(Q_{p}^{\dagger}Q_{h}^{\dagger}+Q_{p}Q_{h})~,
\label{H1}
\end{equation}
which, in the $ph$ symmetric case, coincides with 
Eq. (13) of Ref. \cite{Hirsch} ~.
\subsection{SCRPA equation}
\label{equations}
The SCRPA equation is derived based on the 
RRPA additional and removal 
operators, which have the form
\begin{equation}
    A_{\mu}^{\dagger}=
    \sum_{p}^{\Omega_{p}}{X_{p}^{\mu}}
    \overline{Q}_{p}^{\dagger}-
    \sum_{h}^{\Omega_{h}}{Y_{h}^{\mu}}\overline{Q}_{h}~,
\label{A}
\end{equation}
and
\begin{equation}
    R_{\lambda}^{\dagger}=
    \sum_{h}^{\Omega_{h}}
    {X_{h}^{\lambda}}\overline{Q}_{h}^{\dagger}-
    \sum_{p}^{\Omega_{p}}Y_{p}^{\lambda}\overline{Q}_{p}~,
\label{R}
\end{equation}
respectively~, with the abbreviation
\begin{equation}
\overline{\cal O}_{i}^{\dagger}=\frac{{\cal O}_{i}^{\dagger}}{\sqrt{\langle D_{i}\rangle}}~,
\hspace{5mm}
\overline{\cal O}_{i}=\frac{{\cal O}_{i}}{\sqrt{\langle D_{i}\rangle}}~,\hspace{5mm} 
i=p,~h~
\end{equation}
to denote the renormalized operator of an operator ${\cal O}_{i}^{\dagger}$.
Operator $A_{\mu}^{\dagger}$ transfers the states in a system with $\Omega$
particles to those of a system with $\Omega$+2 particles. 
Operator $R_{\lambda}^{\dagger}$ transfers the states 
of an $\Omega$-particle system to those of a system with $\Omega$-2 particles.
The brackets $\langle\ldots\rangle\equiv\langle\Omega,0|\ldots|\Omega,0\rangle$ 
denote the average over the 
correlated ground state $|\Omega,0\rangle$ of the system with $\Omega$ particles, 
which is defined as the vacuum of operators $A_{\mu}$ and 
$R_{\lambda}$, i.e.
\begin{equation}
A_{\mu}|\Omega,0\rangle=R_{\lambda}|\Omega,0\rangle=0~.
\label{GS}
\end{equation}
Using the exact commutation relation (\ref{[QQ]}) and the definition 
(\ref{GS})
of the RPA ground state $|\Omega,0\rangle$, one can see that the additional and removal operators satisfy the
boson commutation relations in the ground state $|\Omega,0\rangle$
\begin{equation}
    \langle[A_{\mu},A_{\mu'}^{\dagger}]\rangle=\delta_{\mu\mu'}~,\hspace{5mm} 
    \langle[R_{\lambda},R_{\lambda'}^{\dagger}]\rangle=\delta_{\lambda\lambda'}~,
    \label{[AA][RR]}
\end{equation}
if the amplitudes $X$ and $Y$ satisfy the following normalization 
(orthogonality) conditions 
\begin{equation}
\sum_{p}X_{p}^{\mu}X_{p}^{\mu'}-\sum_{h}Y_{h}^{\mu}Y_{h}^{\mu'}=\delta_{\mu\mu'}~,
\hspace{2mm} 
\sum_{h}X_{h}^{\lambda}X_{h}^{\lambda'}-\sum_{p}Y_{p}^{\lambda}Y_{p}^{\lambda'}
    =\delta_{\lambda\lambda'}~,
    \hspace{2mm} 
    \sum_{p}X_{p}^{\mu}Y_{p}^{\lambda}-\sum_{h}X_{h}^{\lambda}Y_{h}^{\mu}=0~,
    \label{norm}
\end{equation}
while the closure relations 
\begin{equation}
\sum_{\mu}X_{p}^{\mu}X_{p'}^{\mu}-\sum_{\lambda}Y_{p}^{\lambda}Y_{p'}^{\lambda}
=\delta_{pp'}~,\hspace{2mm} 
    \sum_{\lambda}X_{h}^{\lambda}X_{h'}^{\lambda}-\sum_{\mu}
    Y_{h}^{\mu}Y_{h'}^{\mu}
    =\delta_{hh'}~,
    \hspace{2mm} 
    \sum_{\lambda}X_{h}^{\lambda}Y_{p}^{\lambda}-\sum_{\mu}X_{p}^{\mu}Y_{h}^{\mu}=0~
    \label{closure}
\end{equation}
guarantee the following inverse transformation of Eqs. (\ref{A}) and (\ref{R})
\[
Q_{p}^{\dagger}=\sqrt{\langle 
D_{p}\rangle}\bigg[\sum_{\mu}X_{p}^{\mu}A_{\mu}^{\dagger}
+\sum_{\lambda}Y_{p}^{\lambda}R_{\lambda}\bigg]~,
\]
\begin{equation}
Q_{h}=\sqrt{\langle 
D_{h}\rangle}\bigg[\sum_{\lambda}X_{h}^{\lambda}R_{\lambda}
+\sum_{\mu}Y_{h}^{\mu}A_{\mu}^{\dagger}\bigg]~.
\label{inverse}
\end{equation}
The SCRPA equation is obtained in a standard way by linearizing the 
equation of motion. The matrix form of the SCRPA equation for the 
additional mode is
\begin{equation}
\left(\begin{array}{cc}{A}&{B}\\
    -{B}&{C}\end{array}
    \right)\left(\begin{array}{c}{X^{\mu}}\\
    {Y^{\mu}}\end{array} \right)=
    E_{\mu}\left(\begin{array}{c}{X^{\mu}}\\
    {Y^{\mu}}\end{array}\right)~,
    \label{SCRPA}
\end{equation}
where the submatrices $A$, $B$, and $C$ were derived in Ref. 
\cite{Hirsch} 
using the definition (\ref{D}) as well as the exact commutation 
relations (\ref{[QQ]}) and (\ref{[MQ]}) as
\[
A_{pp'}=\langle[~\overline{Q}_{p},
[H,\overline{Q}_{p'}^{\dagger}]]\rangle=
\]
\begin{equation}
    2\bigg\{\big[\epsilon(p-\frac{1}{2})+\frac{G}{2}\big]+
\frac{G}{\langle D_{p}\rangle}\big[\sum_{p''}\langle 
Q_{p''}^{\dagger}Q_{p}\rangle-\sum_{h''}\langle 
Q_{p}Q_{h''}\rangle\big]\bigg\}\delta_{pp'}
-G\frac{\langle D_{p}D_{p'}\rangle}{\sqrt{\langle D_{p}\rangle\langle 
D_{p'}\rangle}}~,
\label{Amatrix}
\end{equation}
\begin{equation}
B_{ph}=-\langle[~\overline{Q}_{p},
[H,\overline{Q}_{h}]]\rangle=
G\frac{\langle D_{p}D_{h}\rangle}{\sqrt{\langle D_{p}\rangle\langle 
D_{h}\rangle}}~,
\label{Bmatrix}
\end{equation}
\[
C_{hh'}=-\langle[~\overline{Q}_{h},
[H,\overline{Q}_{h'}^{\dagger}]]\rangle=
\]
\begin{equation}
    -2\bigg\{\big[\epsilon(h-\frac{1}{2})+\frac{G}{2}\big]+
\frac{G}{\langle D_{h}\rangle}\big[\sum_{h''}\langle 
Q_{h}^{\dagger}Q_{h''}\rangle-\sum_{p''}\langle 
Q_{p''}^{\dagger}Q_{h}^{\dagger}\rangle\big]\bigg\}\delta_{hh'}
+G\frac{\langle D_{h}D_{h'}\rangle}{\sqrt{\langle D_{h}\rangle\langle 
D_{h'}\rangle}}~.
\label{Cmatrix}
\end{equation}
The expectation values of the products of two pair operators at the 
right-hand side (rhs) of Eqs. (\ref{Amatrix}) and (\ref{Cmatrix}) are
\begin{equation}
\langle Q_{p}^{\dagger}Q_{p'}\rangle=\langle P_{p}^{\dagger}P_{p'}\rangle=
\sqrt{\langle D_{p}\rangle\langle 
D_{p'}\rangle}\sum_{\lambda}Y_{p}^{\lambda}Y_{p'}^{\lambda}~,
\label{pp}
\end{equation}
\begin{equation}
\langle Q_{p}Q_{h}\rangle=\langle Q_{h}^{\dagger}Q_{p}^{\dagger}\rangle=
-\langle P_{h}^{\dagger}P_{p}\rangle=
-\langle P_{p}^{\dagger}P_{h}\rangle=
\sqrt{\langle D_{p}\rangle\langle 
D_{h}\rangle}\sum_{\lambda}X_{h}^{\lambda}Y_{p}^{\lambda}~,
\label{ph}
\end{equation}
\begin{equation}
\langle Q_{h}^{\dagger}Q_{h'}\rangle=
\sqrt{\langle D_{h}\rangle\langle 
D_{h'}\rangle}\sum_{\mu}Y_{h}^{\mu}Y_{h'}^{\mu}=
\langle P_{h'}^{\dagger}P_{h}\rangle
-\delta_{hh'}\langle 
D_{h}\rangle~,
\label{hh}
\end{equation}
\[
{\rm with}\hspace{3mm} 
\langle P_{h'}^{\dagger}P_{h}\rangle=\sqrt{\langle D_{h}\rangle\langle 
D_{h'}\rangle}\sum_{\lambda}X_{h}^{\lambda}X_{h'}^{\lambda}~.
\]
They were derived using the inverse transformation (\ref{inverse}) 
and the definition of the ground state (\ref{GS}).
The RRPA equation was obtained from Eqs. (\ref{SCRPA}) -- (\ref{Cmatrix}) 
by using the factorization 
\begin{equation}
    \langle D_{i}D_{j}\rangle\simeq\langle D_{i}\rangle\langle 
    D_{j}\rangle~,
    \label{approx}
\end{equation}
and neglecting all the expectation values 
$\langle Q^{\dagger}_{p'}Q_{p}\rangle$, $\langle Q_{p}Q_{h}\rangle$, 
and $\langle Q_{h}^{\dagger}Q_{h'}\rangle$. The RRPA submatrices
have then the form
\begin{equation}
 A_{pp'}^{\rm RRPA}= 2\big[\epsilon(p-\frac{1}{2})+\frac{G}{2}\big]
 \delta_{pp'}
-G\sqrt{\langle D_{p}\rangle\langle D_{p'}\rangle}~,
\label{ARRPA}
\end{equation}
\begin{equation}
B_{ph}^{\rm RRPA}=
G\sqrt{\langle D_{p}\rangle\langle 
D_{h}\rangle}~,
\label{BRRPA}
\end{equation}
\begin{equation}
 C_{hh'}^{\rm RRPA}=  -2\big[\epsilon(h-\frac{1}{2})+\frac{G}{2}\big]\delta_{hh'}
+G\sqrt{\langle D_{h}\rangle\langle D_{h'}\rangle}~.
\label{CRRPA}
\end{equation}
The RPA submatrices are obtained from the RRPA ones by putting 
$D_{p}=D_{h}=$ 1, namely
\begin{equation}
 A_{pp'}^{\rm RPA}= 2\big[\epsilon(p-\frac{1}{2})+\frac{G}{2}\big]
 \delta_{pp'}-G~,
\label{ARPA}
\end{equation}
\begin{equation}
B_{ph}^{\rm RPA}=G~,
\label{BRPA}
\end{equation}
\begin{equation}
 C_{hh'}^{\rm RPA}=  -2\big[\epsilon(h-\frac{1}{2})+\frac{G}{2}\big]\delta_{hh'}
+G~.
\label{CRPA}
\end{equation}
By using definition (\ref{M&Q}), Eqs. (\ref{pp}) -- (\ref{hh}), and 
recalling that
\begin{equation}
\epsilon_{p}-\lambda=[\epsilon(p-\frac{1}{2})+\frac{G}{2}]~,
\label{epsp}
\end{equation}
\begin{equation}
\epsilon_{h}-\lambda=-\epsilon(h-\frac{1}{2})+\frac{G}{2}=
-[\epsilon(h-\frac{1}{2})+\frac{G}{2}]+G~,
\label{epsh}
\end{equation}
one 
can rewrite Eqs. (\ref{Amatrix}) -- (\ref{Cmatrix}) 
in the notations of Ref. \cite{Dukelsky} as
\footnote
{There are several misprints in Eqs. (11) of
Ref. \cite{Dukelsky}, namely the factor 
2 in front of all $\langle N_{i}N_{j}\rangle$ in the numerators of the 
last terms at the rhs of submatrices $\bar{A}$, $\bar{B}$ and 
$\bar{C}$ should be eliminated. 
Also, the factor $1-\langle N_{p'}\rangle$ in the denominator of the last 
term of $\bar{B}_{ph}$ should be replaced with 
$\langle N_{h}\rangle -1$, and the sign ``--'' in front of 
$2\delta_{hh'}(\epsilon_{hh'}-\lambda)$ in the expression of 
$\bar{C}_{hh'}$ should be reversed.}
\[
A_{pp'}=\langle[\overline{P}_{p},[H,\overline{P}_{p'}^{\dagger}]]\rangle=
2\delta_{pp'}(\epsilon_{p'}-\lambda)
\]
\begin{equation}
+
2G\delta_{pp'}\frac{\sum_{p''}\langle 
P_{p''}^{\dagger}P_{p}\rangle+\sum_{h''}\langle 
P_{h''}^{\dagger}P_{p}\rangle}{1-\langle N_{p}\rangle}
-G\frac{\langle(1-N_{p})(1-N_{p'})\rangle}{\sqrt{(1-\langle 
N_{p}\rangle)(1-\langle N_{p'}\rangle)}}~,
\label{Amatrix1}
\end{equation}
\begin{equation}
B_{ph}=\langle[\overline{P}_{p},[H,\overline{P}_{h}^{\dagger}]]\rangle
=G\frac{\langle(1-N_{p})(N_{h}-1)\rangle}
{\sqrt{(1-\langle N_{p}\rangle)(\langle N_{h}\rangle-1)}}~,
\label{Bmatrix1}
\end{equation}
\[
C_{hh'}=-\langle[\overline{P}_{h},[H,\overline{P}_{h'}^{\dagger}]]\rangle
=2\delta_{hh'}(\epsilon_{h'}-\lambda)\]
\begin{equation}
-2G\delta_{hh'}\frac{\sum_{h''}\langle 
P_{h''}^{\dagger}P_{h}\rangle+\sum_{p''}\langle 
P_{p''}^{\dagger}P_{h}\rangle}{\langle N_{h}\rangle-1}
+G\frac{\langle (N_{h}-1)(N_{h'}-1)\rangle}{\sqrt{(\langle 
N_{h}\rangle-1)(\langle N_{h'}\rangle-1)}}~.
\label{Cmatrix1}
\end{equation}
\subsection{Correlation energy}
\label{Ecorr}
The correlation energy $E_{\rm corr}^{\rm SCRPA}$ 
is defined as the difference between the 
energy ${\cal E}^{\Omega}_{0}\equiv\langle H\rangle$ in the ground state 
$|\Omega,0\rangle$ (\ref{GS}) and the Hartree-Fock (HF) energy ${\cal 
E}_{\rm HF}$. 
The former is easily obtained from Eq. (\ref{H1}) while the later is 
$-\epsilon\Omega_{h}^{2}$. The final expression for the correlation 
energy is obtained as
\[
E_{\rm 
corr}^{\rm SCRPA}\equiv\langle H\rangle-{\cal E}_{\rm HF}=
\sum_{p}^{\Omega_{p}}\big[\epsilon(p-\frac{1}{2})+\frac{G}{2}\big]
(1-\langle 
D_{p}\rangle)+\sum_{h}^{\Omega_{h}}\big[\epsilon(h-\frac{1}{2})+
\frac{G}{2}\big](1-\langle 
D_{h}\rangle)
\]
\begin{equation}
-G[\sum_{pp'}\langle Q_{p}^{\dagger}Q_{p'}\rangle+
\sum_{hh'}\langle Q_{h}^{\dagger}Q_{h'}\rangle-2\sum_{ph}\langle 
Q_{p}^{\dagger}Q_{h}^{\dagger}\rangle]~,
\label{EcSCRPA}
\end{equation}
where $\langle Q_{i}^{\dagger}Q_{i}\rangle$ and $\langle 
Q_{p}^{\dagger}Q_{h}^{\dagger}\rangle$ are given by Eqs. (\ref{pp}) -- 
(\ref{hh}), and $D_{i}$ by Eq. (\ref{DpDh}).

For comparison, the
correlation energy within the RRPA is derived here by
approximating the Hamiltonian (\ref{H}) as
\begin{equation}
H\simeq H^{\rm RRPA}={\cal 
E}_{0}^{(\Omega)}({\rm RRPA})+\sum_{\mu}E_{\mu}^{\rm RRPA}A_{\mu}^{\dagger}
A_{\mu}+
\sum_{\lambda}E_{\lambda}^{\rm RRPA}R_{\lambda}^{\dagger}R_{\lambda}~,
\label{H2}
\end{equation}
where ${\cal E}^{(\Omega)}_{0}({\rm RRPA})$ 
is the energy in the 
RRPA ground state, while the eigenvalues $E_{\mu}^{\rm RRPA}$ 
and
$E_{\lambda}^{\rm RRPA}$ are the excitation energies (\ref{Emu}) and 
(\ref{Elam})
of the additional and removal modes, respectively, which are obtained
by solving the RRPA equation, i.e. Eq. (\ref{SCRPA}) 
with submatrices (\ref{ARRPA}) -- (\ref{CRRPA}).
The correlation energy 
$E_{\rm corr}^{\rm RRPA}={\cal E}_{0}^{\Omega}({\rm RRPA})-{\cal E}_{\rm HF}$ 
is obtained by calculating the expectation value of 
Hamiltonian (\ref{H2}) in the HF ground state $|{\rm HF}\rangle$.
By using Eqs. (\ref{A}) 
and (\ref{R}) as well as the definition of $|{\rm HF}\rangle$, 
for which $P_{p}|{\rm HF}\rangle=\langle{\rm HF}|P_{p}^{\dagger}=
P_{h}^{\dagger}|{\rm HF}\rangle=\langle{\rm HF}|P_{h}=$ 0, we
finally obtain
\begin{equation}
{E}_{\rm corr}^{\rm RRPA}={\cal E}^{\Omega}_{0}({\rm RRPA})
-\langle{\rm HF}|H|{\rm HF}\rangle
=-\sum_{\mu}E_{\mu}^{\rm RRPA}
\sum_{h}\frac{({Y}_{h}^{\mu})^{2}}{D_{h}}
-\sum_{\lambda}E_{\lambda}^{\rm RRPA}\sum_{p}\frac{({Y}_{p}^{\lambda})^{2}}{D_{p}}~.
\label{EcRRPA}
\end{equation}
The RPA correlation energy $E_{\rm corr}^{\rm RPA}$ 
is recovered from Eq. (\ref{EcRRPA}) putting $D_{i}=$ 1, namely
\begin{equation}
    {E}_{\rm corr}^{\rm RPA}=-\sum_{\mu}E_{\mu}^{\rm RPA}\sum_{h}
    (Y_{h}^{\mu})^{2}
    -\sum_{\lambda}E_{\lambda}^{\rm RPA}\sum_{p}({Y}_{p}^{\lambda})^{2}~,
    \label{Ecorr1}   
\end{equation}
with $E_{\mu}^{\rm RPA}$ and $E_{\lambda}^{\rm RPA}$ denoting the
RPA excitation energies of the additional and removal modes, 
respectively. 
\section{Particle-number within SCRPA}
\label{violationN}
The SCRPA Eqs. (\ref{SCRPA}) - (\ref{Cmatrix}) 
has $\Omega_{p}$ solutions for the additional mode with
positive eigenvalues 
\begin{equation}
E_{\mu}={\cal E}^{\Omega+2}_{\mu}-
{\cal E}_{0}^{\Omega}>0~,\hspace{5mm} \mu=1,\ldots,\Omega_{p}~,
\label{Emu}
\end{equation}
which are excitation energies 
of the $\Omega+2$ system relative to the 
ground state of the $\Omega$ system. 
Since the SCRPA equation for the removal mode has exactly the same form 
as that of Eqs. 
(\ref{SCRPA}) -- (\ref{Cmatrix}) with the only difference that  
$(X_{p}^{\mu}, Y_{h}^{\mu}, 
E_{\mu})$ should be replaced 
with $(Y_{p}^{\lambda}, X_{h}^{\lambda}, 
-E_{\lambda})$, the $\Omega_{h}$  
negative eigenvalues 
\begin{equation}
{E}_{\lambda}={\cal E}_{\lambda}^{\Omega-2}-{\cal E}_{0}^{\Omega}<0~, 
\hspace{5mm} \lambda=1,\ldots,\Omega_{h}~,
\label{Elam}
\end{equation}
of Eqs. (\ref{SCRPA}) - (\ref{Cmatrix}) have physical meaning as 
excitation energies of the $\Omega-2$ system relative to the ground 
state of $\Omega$ system. 
This set of equations should be solved
self-consistently with the normalization conditions (\ref{norm}) and 
the equations for the factor $\langle D_{p}\rangle$ and $\langle D_{h}\rangle$, 
which represent the ground-state correlations beyond the RPA.
It is clear from Eq. (\ref{D}) that, in the absence of ground-state 
correlations (beyond RPA), the ground state $|\Omega,0\rangle$ becomes the RPA 
ground state $|{\rm RPA}\rangle$, for which
$\langle{\rm RPA}|D_{i}|{\rm RPA}\rangle=
\langle{\rm HF}|D_{i}|
{\rm HF}\rangle=$ 1~($i=p,~h$) due to the QBA, 
where ${\rm HF}$ denotes the HF ground 
state. This
means that $\langle{\rm RPA}|N_{p}|{\rm RPA} \rangle=$ 0 and 
$\langle{\rm RPA}|N_{h}|{\rm RPA}\rangle=$ 2.
In this case, the SCRPA equation reaches its RPA limit with the RPA 
submatrices (\ref{ARPA}) -- (\ref{CRPA}). 
In the general case, $0<\langle D_{i}\rangle<$ 1 ($i=p,~h$) since
$0<\langle N_{p}\rangle<$ 1 and 
$1<\langle N_{h}\rangle<$ 2.

\subsection{Violation of particle number within SCRPA}
In order to derive the equations for the factors $\langle D_{i}\rangle$ 
($i=p,~h$) Refs. \cite{Hirsch,Dukelsky} employed
a procedure similar to the one used in Ref. \cite{CaDaSa} with
the representation
\begin{equation}
    N_{i}=2P_{i}^{\dagger}P_{i}~,
\label{N}
\end{equation}
which becomes exact for the picket-fence model.
Using Eqs. (\ref{D}), (\ref{pp}) and (\ref{hh}), one
finds immediately from the Eq. (\ref{N})
\[
    \langle N_{p}\rangle=
    2\langle D_{p}\rangle\sum_{\lambda}(Y_{p}^{\lambda})^{2}=
    2(1-\langle N_{p}\rangle)\sum_{\lambda}(Y_{p}^{\lambda})^{2}~,
\]
\begin{equation}
    \langle N_{h}\rangle=
    2\langle D_{h}\rangle[1+\sum_{\mu}(Y_{h}^{\mu})^{2}]=
    2(\langle N_{h}\rangle-1)[1+\sum_{\mu}(Y_{h}^{\mu})^{2}]~.
    \label{N1}
\end{equation}
This yields
\[
\langle N_{p}\rangle=1-\langle D_{p}\rangle=
2\langle D_{p}\rangle\sum_{\lambda}(Y_{p}^{\lambda})^{2}~,
\]
\begin{equation}\langle 
N_{h}\rangle=1+\langle D_{h}\rangle=
2[1-\langle D_{h}\rangle\sum_{\mu}(Y_{h}^{\mu})^{2}]~,
\label{NpNh}
\end{equation}
with
\begin{equation}
    \langle D_{p}\rangle=\frac{1}{1+2\sum_{\lambda}(Y_{p}^{\lambda})^{2}}~,
    \hspace{5mm} 
    \langle D_{h}\rangle=\frac{1}{1+2\sum_{\mu}(Y_{h}^{\mu})^{2}}~.
    \label{DpDh}
\end{equation}
This results is a special (degenerated) case of 
the equations for the $pp$ and $hh$ ground-state correlation factors 
$\langle D_{pp'}\rangle$ and $\langle D_{hh'}\rangle$ 
in the general realistic spherical shell-model  basis, which 
is derived here using the general expression of the relation (\ref{N})
in the form
\begin{equation}
N_{j}=\sum_{m}c_{jm}^{\dagger}c_{jm}\rightarrow
\sum_{JMj'} 
P^{\dagger}_{jj'}(JM)P_{jj'}(JM)~,
\label{nj}
\end{equation}
with
\begin{equation}
P_{jj'}^{\dagger}(JM)=\sum_{mm'}\langle jmj'm'|J'M'\rangle
c^{\dagger}_{jm}c^{\dagger}_{j'm'}~,\hspace{5mm} 
P_{jj'}(JM)=[P_{jj'}^{\dagger}(JM)]^{\dagger}~.
\end{equation}
Inserting in the rhs of Eq. (\ref{nj}) 
the general expression for $P_{jj'}^{\dagger}(JM)$ 
\[
    P_{pp'}^{\dagger}(JM)=\sqrt{\langle D_{pp'}\rangle}\bigg(
    \sum_{{\mu}}X_{pp'}^{J{\mu}}
    A_{JM{\mu}}^{\dagger}-\sum_{{\lambda}}
    Y_{pp'}^{J{\lambda}}R_{{JM}{\lambda}}\bigg)~,
\]
\begin{equation}
    P_{hh'}^{\dagger}({JM})=\sqrt{\langle D_{hh'}\rangle}\bigg(
    \sum_{{\lambda}}X_{hh'}^{J\lambda}
    R_{{JM}{\lambda}}^{\dagger}-\sum_{{\mu}}
    Y_{hh'}^{J{\mu}}A_{JM{\mu}}\bigg)~,
\end{equation}
and using Eqs. (\ref{GS}) and (\ref{[AA][RR]}), 
the final equations for $\langle D_{jj'}\rangle\equiv 1-n_{j}-n_{j'}$ 
is obtained in the 
form
\[
    \langle D_{pp'}\rangle
    =1-\sum_{J,\lambda}(2J+1)\sum_{p"}\bigg[
    \langle D_{pp"}\rangle
    \frac{|Y^{J\lambda}_{pp"}|^{2}}{(2j_{p}+1)}
    +\langle D_{p"p'}\rangle\frac{|Y^{J{\lambda}}_{p"p'}|^{2}}
    {(2j_{p'}+1)}\bigg]~,    
\]
\begin{equation}    
    \langle D_{hh'}\rangle=1-\sum_{J,{\mu}}(2J+1)\sum_{h"}\bigg[
    \langle D_{hh"}\rangle\frac{|Y^{J{\mu}}_{hh"}|^{2}}{(2j_{h}+1)}
    +\langle D_{h"h'}\rangle
    \frac{|Y^{J{\mu}}_{h"h'}|^{2}}{(2j_{h'}+1)}\bigg]~,    
\label{Djj'}
\end{equation}
where $j_{p}$ ($j_{h}$) denotes a $p$ ($h$) orbital angular momentum, 
and  
$J$ is the total angular momentum (multipolarity of the excitation). 
Obviously, Eq. (\ref{DpDh}) is recovered from Eq. (\ref{Djj'}) 
in the degenerated case, when $J=j_{p}=j_{h}=0$~, $p=p'$~, and $h=h'$.

Equations (\ref{NpNh}) and (\ref{DpDh}) are the result given by Eq. (13) of Ref. 
\cite{Dukelsky}~\footnote{The index $\lambda$ in the sum at the rhs of 
the expression for $\langle N_{p}\rangle$ in Eq. (13) of Ref. 
\cite{Dukelsky} has been misprinted as $\mu$, although this did not 
affect the results of calculations for the $ph$ symmetric case, for 
which $\mu=\lambda$.}, and Eq. (30) 
of Ref. \cite{Hirsch}. This result for the $pp$ case is similar to what
obtained previously in Ref. \cite{CaDaSa}, but for the $ph$ case, according to 
which
\begin{equation}
    \langle\langle N_{p}\rangle\rangle=
    2\sum_{h\nu}\langle\langle D_{ph}\rangle\rangle(Y_{ph}^{\nu})^{2}~,\hspace{5mm} 
    \langle\langle N_{h}\rangle\rangle=
    2[1-\sum_{p\nu}\langle\langle D_{ph}\rangle\rangle(Y_{ph}^{\nu})^{2}]~,
    \label{Nph}
\end{equation}
where
\begin{equation}
\langle\langle D_{ph}\rangle\rangle=\frac{1}{1+2\sum_{\nu}(Y_{ph}^{\nu})^{2}}~.
\label{Dph}
\end{equation}
Here, to avoid confusion with the notation for the $pp$ case, the double 
brackets $\langle\langle\ldots\rangle\rangle$ are used to denote the
average over the correlated ground state with respect to the
$ph$ renormalized RPA operators.
Except for this formal similarity, the essential difference between the $ph$ and $pp$ cases is that 
Eq. (\ref{Nph}) for 
the $ph$ case always conserves the particle number, while 
Eq. (\ref{NpNh}) for the $pp$ case, in general, does not. Indeed, in the $ph$ case, 
Eq. (\ref{Nph}) gives
$\sum_{p}\langle\langle N_{p}\rangle\rangle\sum_{p}
+\sum_{h}\langle\langle N_{h}\rangle\rangle=2\sum_{h}^{\Omega/2}1=\Omega$ because the 
two sums at the rhs of Eq. (\ref{NpNh}) cancel each other exactly 
and, therefore, independently on how $\langle\langle 
D_{ph}\rangle\rangle$ is estimated.
Meanwhile, in the $pp$ case, 
Eq. (\ref{NpNh}), in general, violates the particle 
number because it gives
\[
\sum_{p}\langle N_{p}\rangle +\sum_{h}\langle N_{h}\rangle
=\Omega+2[\sum_{p}\langle D_{p}\rangle\sum_{\lambda}(Y_{p}^{\lambda})^{2}
-\sum_{h}\langle D_{h}\rangle\sum_{\mu}(Y_{h}^{\mu})^{2}]    
\]
\begin{equation}
=\Omega+2\sum_{\mu}\big[\sum_{p}\langle D_{p}\rangle (X_{p}^{\mu})^{2}
-\sum_{h}\langle D_{h}\rangle (Y_{h}^{\mu})^{2}\big]-
2\sum_{p}\langle D_{p}\rangle
\neq\Omega~,
\label{check1}
\end{equation}
as
\begin{equation}
    \sum_{\mu}\big[\sum_{p}\langle D_{p}\rangle (X_{p}^{\mu})^{2}
    -\sum_{h}\langle D_{h}\rangle (Y_{h}^{\mu})^{2}\big]\neq
    \sum_{p}\langle D_{p}\rangle
    \label{nonorm}
\end{equation}
unless the condition 
$|Y_{p}^{\lambda}|=|Y_{h}^{\mu}|$ is assumed, which means $\langle 
D_{p}\rangle=\langle D_{h}\rangle$. 
This condition is 
satisfied only when the full $ph$ symmetry holds, i.e. 
$\Omega_{p}=\Omega_{h}={\Omega}/2$ for the equidistant spectrum.
In the general $ph$ non-symmetric case, i.e.
when $\Omega_{p}\neq\Omega_{h}\neq\Omega/2$ and/or the spectrum is not 
equidistant, 
there is no normalization condition 
such that (\ref{nonorm}) becomes an equality since this would be 
incompatible with the normalization condition (\ref{norm}) for the SCRPA  
$X_{p}^{\mu}$ and 
$Y_{h}^{\mu}$ amplitudes. 

The measure $\delta\Omega$ 
of particle-number violation can be estimated by expanding $\langle 
D_{i}\rangle$ into
the power series of $(Y_{i}^{\nu})^{2}$. By using the normalization
and closure relations (\ref{norm}) and (\ref{closure}), the lowest 
order of this expansion yields
\[
    |\delta\Omega|\equiv\big|\sum_{p}\langle N_{p}\rangle+\sum_{h}\langle 
    N_{h}\rangle - \Omega\big|=
\big|2\sum_{\mu}\big[\sum_{p}\langle D_{p}\rangle (X_{p}^{\mu})^{2}
-\sum_{h}\langle D_{h}\rangle (Y_{h}^{\mu})^{2}
\big]
-2\sum_{p}\langle D_{p}\rangle\big|
\]
\begin{equation}
    \simeq
\big|2\sum_{\mu}\big\{\sum_{p}[1-2\sum_{\lambda}(Y_{p}^{\lambda})^{2}]
(X_{p}^{\mu})^{2}
-\sum_{h}[1-2\sum_{\mu'}(Y_{h}^{\mu'})^{2}](Y_{h}^{\mu})^{2}
\big\}-2\sum_{p}[1-2\sum_{\lambda}(Y_{p}^{\lambda})^{2}]\big|
\label{dOm}
\end{equation}
\[
=4\big|\sum_{\mu\mu' h}(Y_{h}^{\mu})^{2}(Y_{h}^{\mu'})^{2}
-\sum_{\lambda\lambda' p}(Y_{p}^{\lambda})^{2}(Y_{p}^{\lambda'})^{2}
\big|
\sim {\cal O}(Y^{4})~,
\]
which means that the particle-number violation is expected to be 
small at least within the validity region of RPA, where 
$|Y_{i}^{\nu}|$ are small.
\subsection{Restoration of particle-number conservation within SCRPA}
Equations (\ref{NpNh}) have been derived making use of 
Eq. (\ref{N}), which is compatible only with the exact 
ground state $|\Omega,0\rangle$ (\ref{GS}). 
However,  as has been discussed in details in 
Ref. \cite{Hirsch}, such exact ground state
does not exist within the SCRPA, except for the case with $\Omega=$ 
2, where the SCRPA and exact solutions coincide. Consequently,
the SCRPA formalism still contains some violation of Pauli principle, 
which leads to the particle-number violation in the 
$ph$ non-symmetric case.

In order to restore the particle number within the SCRPA, let us notice 
that the essential point 
of SCRPA and RRPA is the renormalization 
of the operators $Q_{i}^{\dagger}$ and $Q_{i}$ (\ref{M&Q}) 
in such a way that the renormalized operators incorporate the effect 
of Pauli principle due to their fermion structure, 
but behave at the same time like ideal boson operators with respect to 
the expectation value $\langle\ldots\rangle
\equiv\langle\Omega,0|\ldots|\Omega,0\rangle$~\cite{Hara,Rowe}. 
The result of such 
renormalization yields the operators 
$\overline{Q}^{\dagger}_{i}\equiv Q_{i}^{\dagger}/\sqrt{\langle D_{i}\rangle}$ 
and $\overline{Q}_{i}\equiv Q_{i}/\sqrt{\langle D_{i}\rangle}$ 
in Eqs. (\ref{A}) and (\ref{R}), which satisfy the 
following exact relation
\begin{equation}
\langle [\overline{Q}_{i},\overline{Q}_{j}^{\dagger}]\rangle=\delta_{ij}~.
\label{[Qb,Qb]}
\end{equation}
This relation means that, as far as the calculation of expectation 
values is concerned, 
the commutator $[\overline{Q}_{i},\overline{Q}^{\dagger}_{j}]$ can be
safely replaced with its ground-state expectation value 
$\langle [\overline{Q}_{i},\overline{Q}_{j}^{\dagger}]\rangle$, namely
\begin{equation}
    [\overline{Q}_{i},\overline{Q}^{\dagger}_{j}]=
    \langle 
    [\overline{Q}_{i},\overline{Q}_{j}^{\dagger}]\rangle=\delta_{ij}~,
    \label{replace}
\end{equation}
i.e. $\overline{Q}_{i}^{\dagger}$ and $\overline{Q}_{i}$ are now 
considered as ideal boson operators (without fermion structure).
From now on the derivation is proceeded only with expectation 
values using the replacement (\ref{replace}) under the condition
\begin{equation}
\langle A^{\dagger}_{\mu}A_{\mu'}\rangle=
\langle R^{\dagger}_{\lambda}R_{\lambda'}\rangle=0~
\label{condit}
\end{equation}
instead of the vacuum condition (\ref{GS}).\footnote{This situation is somewhat similar to that of 
the statistical formalism, where a quantum mechanical ground state $|0(\beta)\rangle$ 
 ($\beta$ is the inverse temperature) so that $\langle 
 0(\beta)|\hat{\cal O}|0(\beta)\rangle={\rm Tr}\{{\cal O}{\cal D}\}$ 
 (${\cal D}$ 
 is the density operator) does not exist. The expectation value
 $\langle\hat{\cal O}\rangle$ is then replaced with the statistical 
 average over the grand canonical ensemble with ${\cal 
 D}=e^{-\beta(H-\lambda\hat{N})}/{\rm 
 Tr}\{e^{-\beta(H-\lambda\hat{N})}\}$.}
 Using Eq. (\ref{[MQ]}), one can see that the renormalized operators 
 $\overline{Q}_{j}^{\dagger}$ and $\overline{Q}_{j}$ satisfy the following exact commutation 
relations with operators $M_{i}$~:
\begin{equation}
[M_{i},\overline{Q}_{j}^{\dagger}]=2\delta_{ij}\overline{Q}_{i}^{\dagger}~,
\hspace{5mm} [M_{i},\overline{Q}_{j}]=-2\delta_{ij}\overline{Q}_{i}~.
\label{[M,Qb]}
\end{equation}
Since the standard derivation of RRPA equations is based on the algebra 
(\ref{replace}) and (\ref{[M,Qb]}) in terms 
of the boson operators $\overline{Q}_{i}^{\dagger}$ and 
$\overline{Q}_{i}$~\cite{Hara,Rowe,CaDaSa}, in order to derive the equations for the 
renormalization factor $\langle D_{i}\rangle$, 
the fermion operators $M_{i}$ are also bosonized so that Eq. 
(\ref{[M,Qb]}) still remains intact. Such boson representation exists 
and equal to
\begin{equation}
    M_{i}=2\overline{Q}_{i}^{\dagger}\overline{Q}_{i}~,
    \label{N2}
\end{equation}
which fulfills  Eq. (\ref{[M,Qb]}) exactly since
\[
[M_{i},\overline{Q}_{j}^{\dagger}]=2\overline{Q}_{i}^{\dagger}[\overline{Q}_{i},
\overline{Q}_{j}^{\dagger}]=2\delta_{ij}\overline{Q}_{i}^{\dagger}~,
\]
\begin{equation}
    [M_{i},\overline{Q}_{j}]=2[\overline{Q}_{i}^{\dagger},
\overline{Q}_{j}]\overline{Q}_{i}=-2\delta_{ij}\overline{Q}_{i}~,
\label{[M,Qb]1}
\end{equation}
due to Eq. (\ref{replace}). Representation (\ref{N2}) is 
apparently different from Eq. (\ref{N}) since the latter is 
equivalent to
\begin{equation}
M_{i}=2Q_{i}^{\dagger}Q_{i}=2\langle 
D_{i}\rangle\overline{Q}_{i}^{\dagger}\overline{Q}_{i}~
\label{boson1}
\end{equation}
due to definition (\ref{M&Q}). The commutation relations between
Eq. (\ref{boson1}) and operators $\overline{Q}_{i}^{\dagger}$ and 
$\overline{Q}_{i}$ are different from the exact relations 
(\ref{[M,Qb]}) by the factor $\langle D_{i}\rangle$, which causes
the particle-number violation discussed in the preceding section.

Using the representation (\ref{N2}) instead of (\ref{boson1}) together 
with Eqs. (\ref{pp}) and (\ref{hh}) immediately leads to 
\begin{equation}
    \langle N_{p}\rangle=\langle M_{p}\rangle=2\sum_{\lambda}(Y_{p}^{\lambda})^{2}~,
    \hspace{5mm} 
    \langle N_{h}\rangle=2-\langle M_{h}\rangle=2[1-\sum_{\mu}(Y_{h}^{\mu})^{2}]~,
    \label{NpNh2}
\end{equation}
or
\begin{equation}
    \langle D_{p}\rangle=1-2\sum_{\lambda}(Y_{p}^{\lambda})^{2}~,
    \hspace{5mm} 
    \langle D_{h}\rangle=1-2\sum_{\mu}(Y_{h}^{\mu})^{2}~,
    \label{DpDhHara}
\end{equation}
instead of Eqs. (\ref{NpNh}) and (\ref{DpDh}). As has been mentioned 
previously, since 0$\leq\langle 
N_{p}\rangle\leq$ 1 and 1 $\leq\langle 
N_{h}\rangle\leq$ 2, the values of ground-state factors given in Eq. (65) 
should also satisfy $0\leq D_{i}\leq$ 1.
These results are similar to what obtained 
by Hara in Ref. \cite{Hara} and, later by Rowe in Ref. \cite{Rowe} 
but for the $ph$ case. Therefore, the version of 
SCRPA (RRPA), which uses Eqs. (\ref{NpNh2}) and (\ref{DpDhHara}) instead 
of Eqs. (\ref{NpNh}) and (\ref{DpDh}), will be referred to 
as Hara SCRPA (Hara RRPA) hereafter~\footnote{The expressions for the factor $D_{pp'}$ and $D_{hh'}$ 
within the Hara-SCRPA for the general shell-model spherical basis 
are obtained from Eqs. (\ref{Djj'}) putting $D_{jj'}$ 
at the rhs of Eqs. (\ref{Djj'}) equal to 1.}. That Eq. (\ref{NpNh2}) conserves the 
particle number is straightforward, making use of the normalization and closure 
relations 
(\ref{norm}) and (\ref{closure}). Indeed, using Eq. (\ref{NpNh2}) 
instead of Eq. (\ref{NpNh}), one finds in the same way as it was done 
in proving the particle-number conservation within the $pp$ RPA that
\[
    \sum_{p}\langle N_{p}\rangle+\sum_{h}\langle N_{h}\rangle =
    \Omega+2[\sum_{p}\sum_{\lambda}(Y_{p}^{\lambda})^{2}
    -\sum_{h}\sum_{\mu}(Y_{h}^{\mu})^{2}]
\]
\begin{equation}
=
    \Omega+2\big\{\sum_{p}[\sum_{\mu}(X^{\mu}_{p})^{2}-1]
    -\sum_{h}\sum_{\mu}(Y_{h}^{\mu})^{2}\big\}
    \label{check2}
\end{equation}
\[=\Omega+  
    \sum_{\mu}\big[\sum_{p}(X_{p}^{\mu})^{2}-\sum_{h}(Y_{h}^{\mu})^{2}\big]
    -\Omega_{p}
=\Omega+\Omega_{p}-\Omega_{p}=\Omega~.
\]
Therefore, the Hara SCRPA and Hara RRPA always conserve the particle 
number exactly.

\section{Numerical analysis}
\label{numerical}
The calculations were carried out for several values of $\Omega$ 
and $\epsilon=$ 1 MeV within SCRPA and RRPA. 
The most representative case with $\Omega=$ 
10 is selected here for discussion. For simplicity, the 
factorization (\ref{approx}) was used, which has been verified in Ref. 
\cite{Hirsch} to yield excellent results compared with those obtained
when an involved set of 
nonlinear equations for the expectation values $\langle 
M_{i}M_{j}\rangle$ was solved instead. This factorization 
does not affect the discussion regarding the particle-number 
restoration in this work.
\subsection{Degree of particle-number violation within SCRPA}
\begin{figure}                                                             
\includegraphics{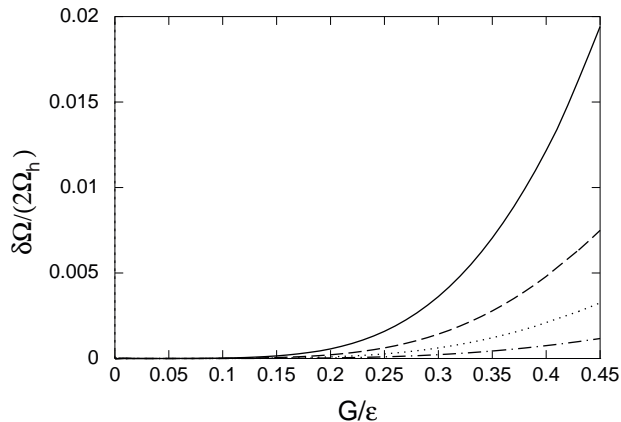}
\caption{\label{dN10}
Particle-number violations within SCRPA for $ph$ non-symmetric cases with 
$\Omega=$ 10 and $\Omega_{h}=$ 1 (solid line), 2 (dashed line), 3 
(dotted line), and 4 (dash-dotted line) as functions of interaction 
parameter $G$ (in units of the level distance $\epsilon$).}
\end{figure}

Shown in Fig. \ref{dN10} is the quantity $\delta\Omega/(2\Omega_{h})$ as a 
function of the interaction parameter $G$ (in units of level distance 
$\epsilon$), which has been obtained within SCRPA for $ph$ non-symmetric cases 
with the number of hole levels $\Omega_{h}=$ 1, 2, 3, and 4.
The particle-number violation increases with $G$ and with the
asymmetry of the $ph$ single-particle space. The strongest violation 
of about 2$\%$ is observed at the strongest asymmetry, i.e. with $\Omega_{h}=1$ 
and $\Omega_{p}=9$ (solid lines), at the largest value of $G=$ 0.45 
MeV shown in the figure. In all other cases plotted on this figure, 
the particle-number violation is smaller than 1$\%$. With increasing 
$\Omega_{h}$, the symmetry is gradually restored, and the 
particle-number violation decreases to reach zero at $\Omega_{h}=$ 5. 
Results of our 
calculations for larger $\Omega$ also show that the particle-number 
violation within SCRPA decreases with increasing the particle number.
\subsection{Correlation, ground-state and excited-state energies}
\begin{figure}                                                             
\includegraphics{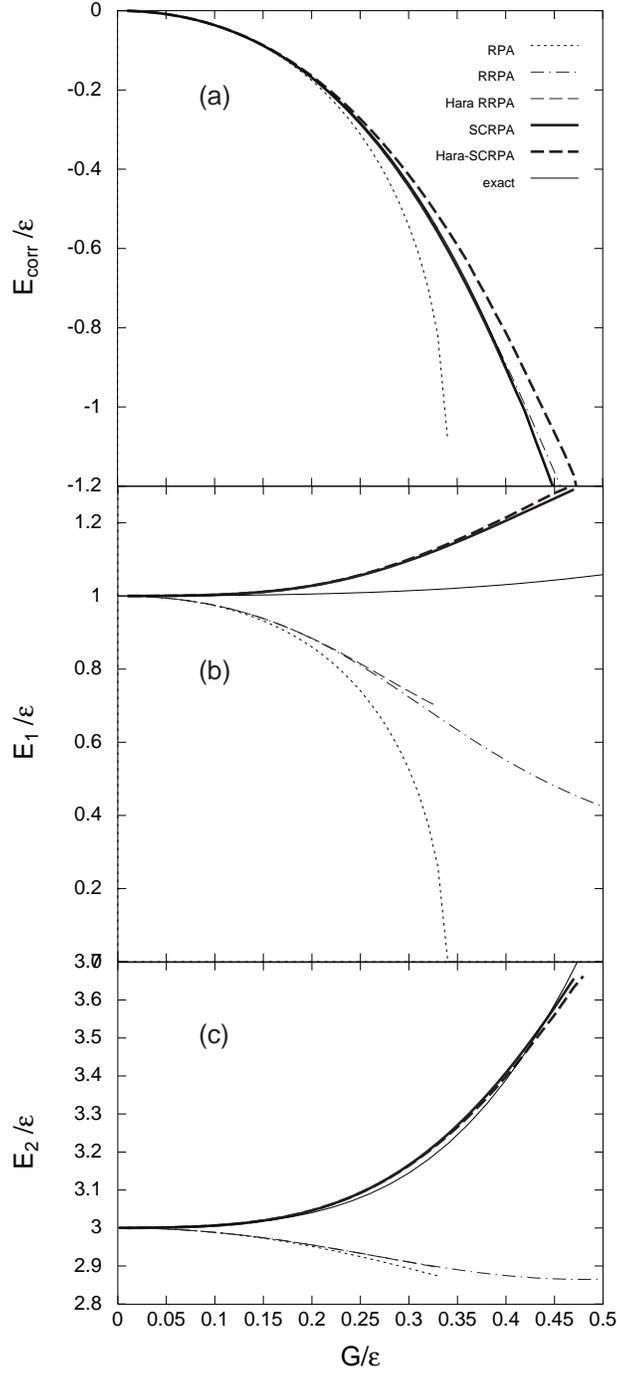}
\caption{\label{EcE1E2}
Correlation energy $E_{\rm corr}$ in the system with $\Omega=10$ 
particles (a), ground-state energy $E_{1}$ (b) 
and first-excited-state energy $E_{2}$ (c) of the system with 
$\Omega+2=$ 12 particles relative to the ground-state of 
$\Omega$-particle system as functions of interaction parameter $G$ 
obtained within RPA (dotted line), RRPA (dash-dotted line), Hara-RRPA 
(thin dashed line), SCRPA (thick solid line), Hara-SCRPA (thick dashed 
line), and exact (thin solid line) calculations.}
\end{figure}
Shown in Fig. \ref{EcE1E2} are the correlations energies of the system 
with $\Omega=10$ particles, as well as the energies $E_{1}$ and 
$E_{2}$ of the ground state and first excited state, respectively, of
the system with $\Omega+2=12$ particles relative to the ground state 
of the $\Omega$-particle system as functions of the interaction 
parameter $G$ (in units of $\epsilon$). They were obtained within the RPA, RRPA, SCRPA, and are 
plotted in comparison with the exact results.
The RRPA gives a quite good description of the correlation energy, 
which practically coincides with that given by the SCRPA and the exact 
result for $G\leq$ 0.45 MeV. However, the RRPA fails badly in 
describing the
the ground-state $E_{1}$ and first-excited-state $E_{2}$ energies of
the $\Omega+2$ system. Here, although the RRPA results do not collapse
at $G_{\rm cr}\simeq$ 0.34 MeV as the RPA results do, they decrease 
monotonously, while the exact results as well as those given by the 
SCRPA increase with increasing $G$ (Cf. Refs. \cite{Dukelsky,Hirsch}).
The results obtained within the Hara-SCRPA are close to those given by 
the RRPA, but fail to converge in this 
model at $G>G_{\rm cr}$.
The Hara-SCRRPA, which conserves the particle-number exactly in $ph$ 
non-symmetric cases, offers very close results to those given by the 
SCRPA for the $E_{1}$ and $E_{2}$ energies within the whole interval 
of values for $G$ under consideration. However, the correlation energy
$E_{\rm corr}$ 
obtained within this number-conserving version of SCRPA is slightly 
larger than the exact result, and the discrepancy is 
clearly visible already starting from $G\geq$ 0.3 MeV.

\begin{figure}                                                             
\includegraphics{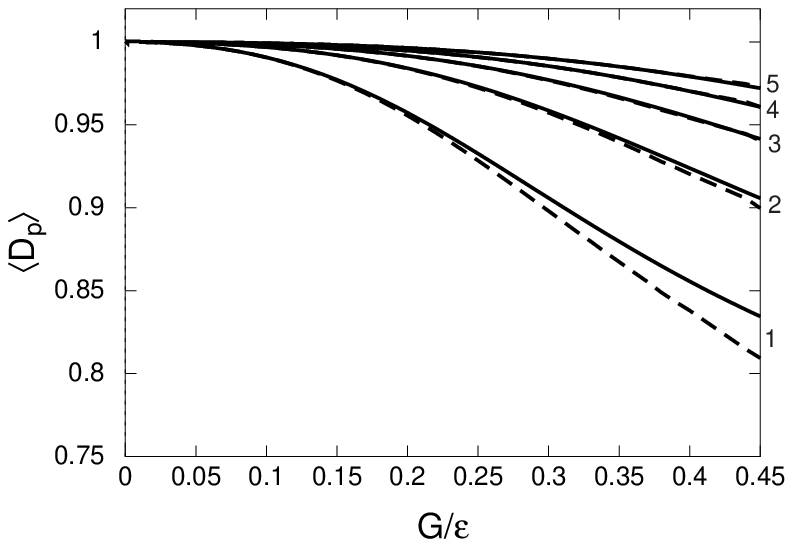}
\caption{\label{Dp}
Ground-state correlation factor $\langle D_{p}\rangle$ for particle levels as 
functions of interaction parameter $G$. The solid and dashed lines 
denote results obtained by using Eqs. (\ref{DpDh}) and 
(\ref{DpDhHara}), respectively. The numbers next to the lines  
numerate the particle level, to which the lines correspond.}
\end{figure}
In general, the feature depicted in Fig. \ref{EcE1E2} 
is similar to that of the $ph$ case considered in 
Ref. \cite{CaDaSa}, where the solution obtained by using Eq. (\ref{Dph})
approaches the exact solution at large $G$, while the one offered by 
the Hara approach fails to describe it, never approaching zero. 
This result comes from the overestimation of ground-state correlations beyond 
RPA within the Hara approach, which can be clearly seen by examining 
the ground-state correlation factors $\langle D_{i}\rangle$ and/or the occupation 
number $N_{i}$. The factor $\langle D_{p}\rangle$, 
which is the same as $\langle D_{h}\rangle$ for 
the symmetric case of the picket-fence model, is shown in Fig. \ref{Dp}.
This factor decreases from 1 with increasing $G$, approaching zero as 
$G\rightarrow\infty$. The deviation from 1 is stronger at the level 
closer to the Fermi one. The difference between the results 
obtained by using Eqs (\ref{Dph}) (the SCRPA) and 
(\ref{DpDhHara}) (the Hara-SCRPA)   
is strongest for the lowest particle level, in which the Hara-SCRPA 
gives stronger ground-state correlations beyond RPA. It also 
increases with increasing $G$ in line with the results obtained for 
the $ph$ case in Ref. \cite{CaDaSa}.
This also explains the larger 
discrepancy between the two approaches in the description of 
correlation energy $E_{\rm corr}$, while the differences in the 
energies $E_{1}$ of the 
ground states, and $E_{2}$ of the first excited states are relatively smaller.

\begin{figure}                                                             
\includegraphics{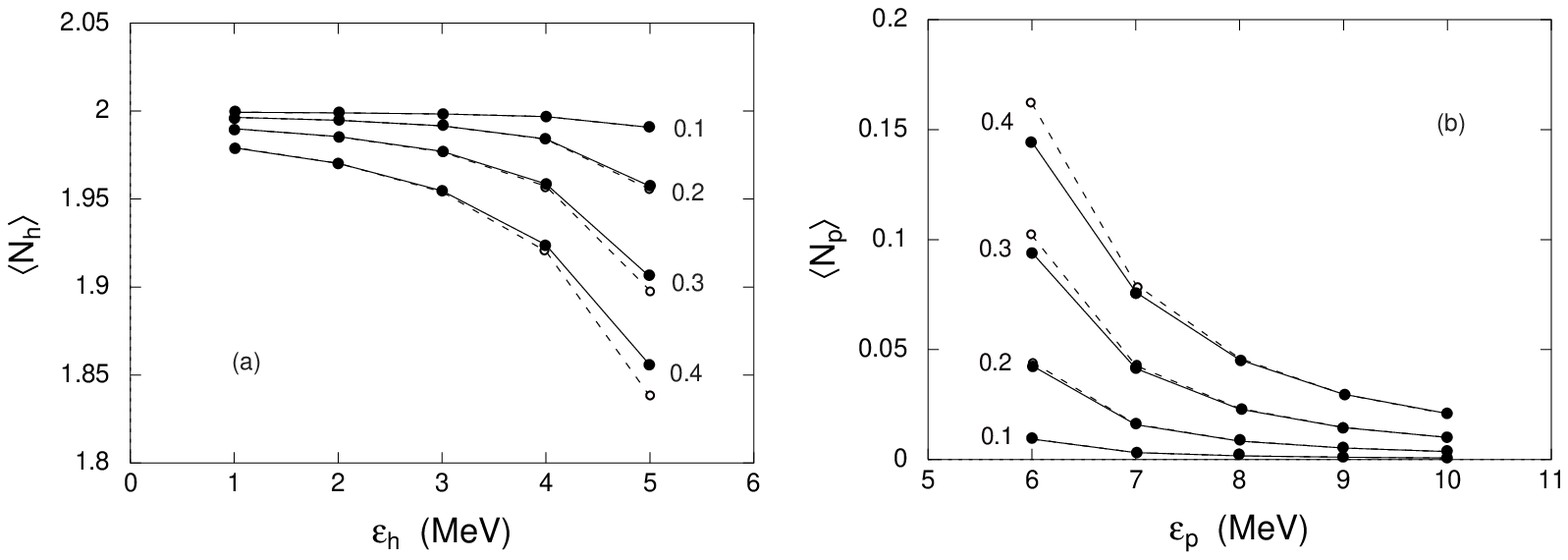}
\caption{\label{N10}
Occupation numbers $\langle N_{h}\rangle$ for hole levels (a) 
and $\langle N_{p}\rangle$ for particle 
levels as functions of single-particle energies $\epsilon_{i}$ at 
several values of $G/\epsilon$ indicated as the numbers next to the 
circles. The full and open circles 
denote results obtained by using Eqs. (\ref{DpDh}) and 
(\ref{DpDhHara}), respectively. The lines are drawn to guide the eyes.}
\end{figure}
The behavior of $\langle D_{i}\rangle$ leads to the change of the occupation number $\langle N_{p}\rangle$ and $\langle N_{h}\rangle$ 
as shown in Fig. \ref{N10}. As 
small $G$, the function $N_{i}$ approaches the stair case with 
$\langle N_{p}\rangle\simeq$ 0 
and $\langle N_{h}\rangle\simeq$ 2 for all $\epsilon_{i}$. As $G$ 
increases, the deviation from the stair case becomes more and more 
evident with the decrease of $\langle N_{h}\rangle$ 
from 2 and increase of $\langle N_{p}\rangle$ 
from 0. At $G/\epsilon=$ 0.4, e.g, $\langle N_{h}\rangle$ 
becomes 1.85 and $\langle N_{p}\rangle$ 
reaches 0.4 for levels closest to the Fermi one.
The deviation caused by the Hara-SCRPA is always stronger than that
given by the SCRPA. Since $\langle D_{i}\rangle\rightarrow$ 0 as 
$G\rightarrow\infty$ we have from Eq. (\ref{DpDhHara}) (the 
Hara-SCRPA) the sum $\sum_{\lambda}(Y_{p}^{\lambda})^{2}=
\sum_{\mu}(Y_{h}^{\mu})^{2}\rightarrow 1/2$, which make 
$N_{p}=N_{h}\rightarrow$ 1. For the SCRPA at the value 
$\sum_{\lambda}(Y_{p}^{\lambda})^{2}=
\sum_{\mu}(Y_{h}^{\mu})^{2}=1/2$ one obtains $\langle D_{p}\rangle=
\langle D_{h}\rangle=1/2$, 
which lead to $\langle N_{p}\rangle=1/2$ and $\langle N_{h}\rangle=3/2$.
Again, this shows that ground-state correlations beyond RPA are 
stronger within Hara-SCRPA than within SCRPA.
\begin{table}
\caption{Correlation energy $E_{\rm corr}$, ground-state energy 
$E_{1}$, and first-excited-state energy $E_{2}$ obtained within the ``parametrized'' Hara-SCRPA 
using Eq. (\ref{para}) with $\alpha=$ 1.9 (a) in comparison with those given by the SCRPA 
(b) for $\Omega=$ 10. All the values are given in MeV. \label{tab}}
\vspace{2mm}

\begin{tabular}{|c|cc|cc|cc|}
    \tableline
    &\multicolumn{2}{c|}{~~~~~~~$E_{\rm corr}$~~~~~~}
    &\multicolumn{2}{c|}{~~~~~~~$E_{1}$~~~~~~}
    &\multicolumn{2}{c|}{~~~~~~~~$E_{2}$~~~~~}\\
\tableline
~~~~$G$~~~~&~~~~~~~~~ a~~~~~~~~~ &~~~~~~~~~ b~~~~~~~~~~ & 
~~~~~~ a~~~~~~ &~~~~~~ b~~~~~~ & 
~~~~~~ a~~~~~~ &~~~~~~ b~~~~~~ \\
\tableline
0.01 & $-0.3265\times10^{-3}$ & $-0.3265\times10^{-3}$ 
& 1.0000 & 1.0000 
& 3.0000 & 3.0000 \\
0.05 & $-0.9018\times10^{-2}$ & $-0.8563\times10^{-2}$ 
& 1.0005 & 1.0005 
& 3.0001 & 3.0001 \\
0.10 & $-0.3850\times10^{-1}$ & $-0.3652\times10^{-1}$ 
& 1.0032 & 1.0033 
& 3.0063 & 3.0063 \\
0.15 & $-0.9263\times10^{-1}$ & $-0.8808\times10^{-1}$ 
& 1.0111 & 1.0112 
& 3.0194 & 3.0194 \\
0.20 & $-0.1760$ & $-0.1686$ 
& 1.0278 & 1.0279 
& 3.0811 & 3.0460 \\
0.25 & $-0.2930$ & $-0.2846$ 
& 1.0566 & 1.0563 
& 3.0922 & 3.0926 \\
0.30 & $-0.4470$ & $-0.4424$ 
& 1.0985 & 1.0970
& 3.1645 & 3.1657 \\
0.35 & $-0.6401$ & $-0.6475$ 
& 1.1514& 1.1481
& 3.2670 & 3.2703 \\
0.40 & $-0.8823$ & $-0.9023$ 
& 1.2118& 1.2058
& 3.4011 & 3.4087 \\
0.45 & $-1.1608$ & $-1.2094$ 
& 1.2755& 1.2667
& 3.5654 & 3.5804 \\
\tableline
\end{tabular}
\end{table}

The exaggeration of the
ground-state correlations beyond RPA within 
the renormalization procedure, which leads to
the number-conserving (Hara) type expressions (\ref{DpDhHara}) 
was pointed out before by Rowe in Ref. \cite{Rowe}, 
where, by using the number-operator 
method to insert the number operator twice at the center of 
$\langle N_{i}\rangle$, he found that the $ph$ ground-state correlation factor 
$\langle\langle D_{ph}\rangle\rangle$ became $\sum_{\nu}|Y_{ph}^{\nu}|^{2}$ instead of 
$2\sum_{\nu}|Y_{ph}^{\nu}|^{2}$. The result of an infinite expansion
by inserting repeatedly the number operator at the center of 
$\langle N_{i}\rangle$ is not available for $pp$RPA at this stage. 
However, the observation by Rowe suggested that the real
$\langle D_{p}\rangle$ and $\langle D_{h}\rangle$ might be closer to 1 than those given by Eqs. 
(\ref{DpDhHara}).
Therefore, a test was also carried out here by parametrizing 
$\langle D_{p}\rangle$ and $\langle D_{h}\rangle$ within the Hara-SCRPA
to see if it is possible to achieve results as good as those given by SCRPA for 
all three quantities $E_{\rm corr}$, $E_{1}$ and $E_{2}$. For this test, we 
used
\begin{equation}
\langle\widetilde{D}_{p}\rangle=1-\alpha\sum_{\lambda}(Y_{p}^{\lambda})^{2}~,
\hspace{5mm} 
\langle\widetilde{D}_{h}\rangle=1-\alpha\sum_{\mu}(Y_{h}^{\mu})^{2}~,\hspace{5mm} 
\alpha<2~,
\label{para}
\end{equation}
instead of Eqs. (\ref{DpDhHara}) and repeated the calculations.
The result of this test shows that 
the values $E_{\rm corr}$, $E_{1}$, and $E_{2}$ 
obtained within the SCRPA can 
be fitted simultaneously rather well 
within such ``parametrized'' Hara-SCRPA with the parameter 
$\alpha\simeq$ 1.9~. These results are shown in Table \ref{tab} in 
comparison with the SCRPA ones. Such ``parametrized'' 
Hara-SCRPA also conserves exactly the particle number in the $ph$ non-symmetric case
as the Hara-SCRPA does.
\section{Conclusions}
The present work shows that the SCRPA
violates the particle number in the $ph$ non-symmetric case if the
occupation numbers are calculated according to Eq. (\ref{NpNh}) for 
the picket-fence model, which is a limit of Eq. (\ref{Djj'}) for the general 
shell-model spherical basis. Within the $ph$ non-symmetric 
picket-fence model this particle-number violation increases with 
the asymmetry and interaction strength $G$, but it decreases with 
increasing the particle number. However, within the interval of values 
for $G$ under consideration ($G\leq$ 0.5 MeV), we also found that the
particle-number violation reaches at most around 0.2$\%$ for the most 
asymmetric case with the level number $\Omega=10$, where 
the number of hole levels $\Omega_{h}=1$, and 
number of particle levels $\Omega_{p}=9$ . 
In all other less asymmetric cases this violation is smaller than 
0.1$\%$.  

In order to maintain the exact particle-number conservation within the SCRPA, 
a renormalization was proposed, which represents
the number operator in terms of
the product of renormalized pairing operators. As a result, a 
number-conserving SCRPA was derived, which is called Hara-SCRPA as it 
has the equations for the occupation numbers similar to what obtained 
in the pioneering works by Hara et al. but for $ph$ case~\cite{Hara}.
The results of numerical calculations show that the Hara-SCRPA yields the ground 
state energy and energy of the first excited state of the $\Omega+2$ 
system very close to the corresponding values obtained 
within the SCRPA. However, the correlation energy, which the 
Hara-SCRPA offers, is slightly larger than that obtained within SCRPA.

The results of the present study also indicate that, in realistic 
calculations using non-symmetric single-particle spectra within RPA, 
in particular for light systems, 
one should carefully examine the violation of Pauli principle to see
if it is important to include the ground-state correlations beyond RPA.
As a matter of fact, the preliminary results of RRPA calculations, 
which were carried out recently for $^{12,14}$Be using the Gogny
interaction~\cite{VMD}, have shown that ground-state correlations beyond RPA 
increased the correlation energy by 20 -- 24$\%$ compared to the RPA 
results. This shifted up the ground state energy by 13$\%$ 
for $^{12}$Be and 48$\%$ for $^{14}$Be. At the same time the particle-number 
violation within the RRPA due to the use of Eq. (\ref{Djj'}) did 
not exceed 0.2$\%$. In this case the SCRPA can be still well justified, and 
has the advantage over the Hara-SCRPA as the former offers a better 
description of the correlation energy. 

In the cases 
where the particle-number 
violation cannot be neglected (e.g. $>1\%$) in calculations with 
realistic spectra and interactions
a number-conserving approach like the Hara-SCRPA proposed in the 
present work might
have to be used instead of the SCRPA. However,
the improvement of the correlation energy in 
this case cannot be achieved by simply renormalizing RPA as has 
been done in the approaches under discussion. 
The test of parametrizing the Hara-SCRPA to yield 
all the three energies $E_{\rm corr}$, $E_{1}$, and $E_{2}$ close to the values 
given by the SCRPA suggests that higher-order correlations may have to 
be included in order to reproduce all these three quantities within a
number-conserving SCRPA. This indicates coupling to 
configurations more complicated than the $ph$,  
$pp$, or $hh$ ones should also be taken into account. 
\acknowledgments
The numerical calculations were carried out using the {\scriptsize 
FOTRAN IMSL} Library 3.0 by Visual Numerics on the Alpha Server 800 
5/500 at the Division of Computer and Information of RIKEN.
The author is grateful to Michelangelo Sambataro for fruitful 
discussions, valuable comments and assistance regarding this work.


\end{document}